\documentclass[12pt,letterpaper]{article}

\usepackage[includeheadfoot,
            marginratio={1:1,2:3}, 
            width=412pt, 
            height=688pt,]{geometry}

\usepackage{amsmath}
\usepackage{amsfonts}
\usepackage{amssymb}
\usepackage{graphicx}
\usepackage{cite}
\usepackage{ulem}
\usepackage{dsfont}
\usepackage{longtable}
\usepackage{afterpage}

\newcommand{\eq}[1]{\begin{equation}
                     \begin{split} #1 \end{split}
                     \end{equation}}

\allowdisplaybreaks[2]
\numberwithin{equation}{section}

\begin{document}

\normalem
\vspace*{-1.5cm}
\begin{flushright}
  {\small
  MPP-2017-80 
  }
\end{flushright}

\vspace{1.5cm}

\begin{center}
  {\LARGE
   ${\cal W}$ algebras are  L$_\infty$ algebras
}
\vspace{0.4cm}

\end{center}

\vspace{0.35cm}
\begin{center}
  Ralph Blumenhagen, Michael Fuchs, Matthias Traube
\end{center}

\vspace{0.1cm}
\begin{center} 
\emph{
Max-Planck-Institut f\"ur Physik (Werner-Heisenberg-Institut), \\ 
F\"ohringer Ring 6,  80805 M\"unchen, Germany 
} 
\end{center} 

\vspace{1cm}


\begin{abstract}
\noindent
It is shown that the closure of the infinitesimal symmetry
transformations underlying classical ${\cal W}$ algebras give rise to 
L$_\infty$ algebras with in general field dependent gauge parameters.
Therefore, the class of well understood  ${\cal W}$ algebras 
provides highly non-trivial examples of such strong homotopy
Lie algebras. We develop the general formalism for this
correspondence and apply it explicitly to the classical ${\cal W}_3$ algebra.
\end{abstract}


\clearpage


\section{Introduction}

Motivated by bosonic closed string field theory \cite{Zwiebach:1992ie}, the structure of
L$_\infty$ algebras was introduced. In the more
mathematical context, these algebras are also called strong homotopy
Lie algebras \cite{Lada:1992wc}. These are
generalizations of Lie algebras and are expected to be closely related 
to the symmetries of classical field theories. The novel feature is
that these algebras do 
contain higher order products as well as generalized Jacobi-identities
among them. Such a structure also appeared in the early work on higher
spin theories in \cite{Berends:1984rq} and also made its appearance in
more recent studies of the Courant bracket \cite{Roytenberg:1998vn} and the C-bracket
\cite{Deser:2016qkw} appearing
in generalized geometry and double field theory, respectively.

Motivated by the study of field theory truncations of string field
theory \cite{Sen:2016qap}, the authors of \cite{Hohm:2017pnh} argued  that the symmetry and the action of any consistent perturbative gauge symmetry is controlled by an L$_\infty$ algebra.
For Chern-Simons and Yang-Mills gauge theories as well as for double field theory the symmetries and equations of motion could be expressed in terms of an  L$_\infty$ structure.

While reading this paper and the early work on higher spin algebras
\cite{Berends:1984rq} (see \cite{Fulp:2002kk} for a more mathematical exposition), as physicists we wondered whether there exist non-trivial
examples that really admit field dependent gauge parameters.
Taking into account that the higher spin algebras  in three dimensions are
holographically dual to two-dimensional conformal field theories with
extended symmetry \cite{Henneaux:2010xg,Campoleoni:2010zq,Gaberdiel:2010pz}, it is natural to contemplate the idea that 
${\cal W}$ algebras might be somehow related to L$_\infty$ algebras.
These ${\cal W}$ algebras have the feature that the commutators of the
generators do not simple close linearly among themselves but involve
also (normal ordered) products of the elementary fields.

Taking into account that the general expectation from  \cite{Zwiebach:1992ie} is, that only
classical symmetries are directly related to  L$_\infty$ algebras, 
in this letter we intend to convey the observation that indeed
classical ${\cal W}$ algebras
provide a large class of highly non-trivial L$_\infty$ algebras
(see e.g. \cite{Bering} for some other concrete examples).
We can give a very generic definition of the higher products between
gauge parameters and fields that are concentrated in the first two
graded vector spaces $X_0$ and $X_{-1}$. In this CFT context, we
recover the general result of
\cite{Berends:1984rq,Burgers:diss,Fulp:2002kk,Hohm:2017pnh}, that the
higher order L$_\infty$ relations are satisfied if and only if the
gauge algebra closes and the Jacobi-identity of for three gauge variations vanishes. With this insight, one can turn the logic around
and {\it bootstrap} the form of the classical ${\cal W}$ algebras by implementing
the L$_\infty$  relations.

First we review some of the for this letter relevant aspects of
L$_\infty$ algebras and ${\cal W}$ algebras. Then we provide the
general
prescription of how a classical ${\cal W}$ algebra induces higher
products and their higher order relations. Some of the ingredients 
can be given in quite general terms and  depend only on the
conformal dimensions of the fields involved. What is left open are some
algebra dependent structure constants. For the concrete example
of the classical ${\cal W}_3$ algebra, we evaluate explicitly the 
 L$_\infty$  relations, which indeed uniquely fix the remaining structure
 constants to precisely the values expected from the closure of the 
${\cal W}_3$ algebra. 


\section{Preliminaries}

In this section we briefly review the basic structures of an
L$_{\infty}$ algebra and of ${\cal W}$ algebras. To keep the
presentation as focused as possible, we will only
introduce those ingredients that are needed for  this letter.


\subsection{Basics of L$_{\infty}$ algebras}

Let us first recall the definition  of an  L$_\infty$ algebra in the
so-called $\ell$-picture. One has a graded vector space 
${X}=\bigoplus_n {X}_n$, where $X_n$ is said to have
degree $n$. In addition there are multi-linear products
$\ell_n(x_1,\ldots,x_n)$ that have degree ${\rm deg}(\ell_n)=n-2$
so that
\eq{
      {\rm deg}\big( \ell_n(x_1,\ldots,x_n)\big)=n-2+\sum_{i=1}^n  {\rm deg}(x_i)\,.
}
Note that $\ell_2$ does not change the degree, while $\ell_1$
decreases it by one.
The products are graded commutative, i.e.
\eq{
            \ell_n(x_{\sigma(1)},\ldots,x_{\sigma(n)})=(-1)^\sigma
            \epsilon(\sigma;x) \, \ell_n(x_{1},\ldots,x_n) \, , 
}
where $(-1)^\sigma$  is just the sign of the permutation $\sigma$ and
the Koszul sign $\epsilon(\sigma;x)$  is defined by considering a graded commutative algebra
$\Lambda(x_1,x_2,\ldots)$ with $x_i\wedge x_j =(-1)^{x_i x_j} x_j\wedge
x_i$ and reading the sign from
\eq{
           x_1\wedge\ldots\wedge x_k=\epsilon(\sigma;x)\,
           x_{\sigma(1)}\wedge \ldots\wedge x_{\sigma(n)}\,.
}
Here $x_i$ in the exponent means ${\rm deg}(x_i)$.
Thus,  e.g. one has
\eq{ 
\label{permuting}
\ell_2 (x_1,x_2) = (-1)^{1+ x_1 x_2} \ell_2 (x_2,x_1)\,.
}
The defining relations of L$_\infty$, labelled by $n$, are  then
\eq{
\label{linftyrels}
{\cal J}_n(x_1,\ldots, x_n):=\sum_{i + j = n + 1 } &(-1)^{i(j-1)} \sum_\sigma (-1)^\sigma
            \epsilon(\sigma;x) \; \\
 &\ell_j \big( \;
\ell_i (x_{\sigma(1)}\; , \dots , x_{\sigma(i)} )\, , x_{\sigma(i+1)} , \dots ,
x_{\sigma(n)} \big) = 0 \,.
}
Here the permutations are restricted to the ones with
\eq{ 
\label{restrictiononpermutation}
\sigma(1) < \cdots < \sigma(i) , \qquad \sigma(i+1) < \cdots < \sigma(n)\,.
}

The L$_\infty$ algebras of interest in this letter are concentrated in
the first two vector spaces $X_0$ and $X_{-1}$, where $X_0$ will contain
gauge parameters $\varepsilon$ and $X_{-1}$ the basic fields $ W$.
In this case, only the $n$-products 
\eq{
\label{higherells}
\ell_n (\varepsilon,W^{n-1})\,,\qquad \ell_n (\varepsilon_1,\varepsilon_2,W^{n-2})
}
can be non-trivial. 

Moreover,   the order $n$ L$_\infty$ relation with $m$ gauge
parameters and $n-m$ fields will be of degree $m-3$.
Therefore, only those relations with $m=2,3$ gauge parameters are
non-trivial. In particular ${\cal J}_1=\ell_1 \ell_1=0$ is trivially satisfied.
For this letter all n-products for $n\ge 4$ vanish so that
the non-trivial relations  are 
${\cal J}_2(\varepsilon_1,\varepsilon_2)=0$, ${\cal
  J}_3(\varepsilon_1,\varepsilon_2, x)=0$,
${\cal J}_4(\varepsilon_1,\varepsilon_2,  W,x)=0$, ${\cal
  J}_5(\varepsilon_1,\varepsilon_2,  W,W,x)=0$  with $x\in\{\varepsilon_3, W\}$.
Their schematic form is 
\eq{ 
&{\cal J}_2 = \ell_1 \ell_2 - \ell_2 \ell_1 \,, \qquad {\cal L}_3 =
\ell_1 \ell_3 + \ell_2 \ell_2 + \ell_3 \ell_1 \,, \\[0.1cm]
& {\cal J}_4 = \ell_1 \ell_4 - \ell_2 \ell_3 + \ell_3 \ell_2 - \ell_4 \ell_1 \, , \\[0.1cm]
& {\cal J}_5 = \ell_1 \ell_5 + \ell_2 \ell_4 + \ell_3 \ell_3 + \ell_4 \ell_2 + \ell_5 \ell_1 \, , 
}
where the sign reflects the factor $(-1)^{i(j-1)}$ in \eqref{linftyrels}.
In our case we have $\ell_4=0$, hence the explicit form of these
relations reads
\eq{
\label{ininftyrel}
        \ell_1\big(\ell_2(\varepsilon_1,\varepsilon_2)\big) = \ell_2\big(\ell_1(\varepsilon_1\big),\varepsilon_2)+\ell_2\big(\varepsilon_1,\ell_1(\varepsilon_2)\big)\,,
}
meaning that  $\ell_1$ acts as a derivative satisfying the Leibniz rule, and 
\begin{eqnarray}    \label{ininftyrel2}
       &&0= \phantom{+} \ell_1\big(\ell_3(\varepsilon_1,\varepsilon_2,x)\big) \nonumber\\
     &&\phantom{0=}+\ell_2\big(\ell_2(\varepsilon_1,\varepsilon_2),x\big)+
     \ell_2\big(\ell_2(\varepsilon_2,x),\varepsilon_1\big)+
     \ell_2\big(\ell_2(x,\varepsilon_1),\varepsilon_2\big) \nonumber\\
     &&\phantom{0=}+\ell_3\big(\ell_1(\varepsilon_1),\varepsilon_2,x\big)+\ell_3\big(\varepsilon_1,\ell_1(\varepsilon_2),x\big)
+\ell_3\big(\varepsilon_1,\varepsilon_2,\ell_1(x)\big)\,, \nonumber\\[0.2cm]
&&0=-\ell_2\big(\ell_3(\varepsilon_1,\varepsilon_2,W ),x\big)+(-1)^{x}
\ell_2\big(\ell_3(\varepsilon_1,\varepsilon_2,x),W \big)\\
&&\phantom{0=}+\ell_2\big(\varepsilon_2,\ell_3(\varepsilon_1,W
,x)\big)-\ell_2\big(\varepsilon_1,\ell_3(\varepsilon_2,W ,x)\big) \nonumber\\[0.1cm]
&&\phantom{0=}+\ell_3\big(\ell_2(\varepsilon_1,\varepsilon_2),W ,x\big)-\ell_3\big(\ell_2(\varepsilon_1,W ),\varepsilon_2,x\big)
+(-1)^{x}
\ell_3\big(\ell_2(\varepsilon_1,x),\varepsilon_2,W \big) \nonumber\\
&&\phantom{0=}+\ell_3\big(\ell_2(\varepsilon_2,W ),\varepsilon_1,x\big)
-(-1)^{x}
\ell_3\big(\ell_2(\varepsilon_2,x),\varepsilon_1,W\big)+\ell_3\big(\ell_2(W
,x),\varepsilon_1,\varepsilon_2\big)\,,\nonumber
\end{eqnarray}       
showing that the usual Jacobi-identity
for $\ell_2$ does receive some correction terms that are $\ell_1$-derivatives. Here we did not spell out the ${\cal J}_5$ relation, since it will turn out to be trivially fulfilled in our case.


\subsection{Basics of ${\cal W}$ algebras}

Let us review just a few facts about ${\cal W}$ algebras, focusing on
the first non-trivial example of the ${\cal W}_3$ algebra. 
For more details on ${\cal W}$ algebras, we refer the reader to the
literature as e.g. the collection of early papers \cite{Bouwknegt:1995ag}.
${\cal W}$ algebras arise as extended symmetry algebras of two-dimensional
conformal field theories. The minimal symmetry algebra is the
Virasoro algebra containing just the generators of infinitesimal
conformal transformations.

It was shown in \cite{Zamolodchikov:1985wn} that one can extend the Virasoro algebra by a
primary field $W(z)$ of conformal dimension three and still get an
algebra that closes and satisfies the Jacobi-identity for the bracket.
The new aspect is that the algebra does not close in the set of
generators themselves but also involves (normal ordered) products of
the latter. In this letter, for concreteness we will focus on the
${\cal W}_3$ algebra
which has two generators $ \mathbf W = \{T,W\}$ of conformal dimension
2 and 3. 
Expanding
\eq{ 
T(z)=\sum_{n\in \mathbb Z} L_n \,z^{-n-2}\,,\qquad 
W(z)=\sum_{n\in \mathbb
  Z} W_n\, z^{-n-3}
} 
defines the modes of the two fields. Their quantum commutator algebra reads
\eq{
\label{w3quant}
                [L_m,L_n]&={c\over 2} {m+1\choose 3} \delta_{m+n, 0} +(m-n) L_{m+n}\,, \\[0.1cm]
               [L_m,W_n]&=(2m-n) W_{m+n}\,, \\[0.1cm]
             [W_m,W_n]&={c\over 3} {m+2\choose 5}\delta_{m+n, 0}+ {\alpha\over 60} (m-n)(2m^2+2n^2-mn-8) L_{m+n}\\
&\phantom{=}+{\beta^{\rm qu}\over 2} (m-n) \Lambda^{\rm qu}_{m+n}
}
with the normal ordered product  $\Lambda^{\rm qu}=N(TT)-{3\over 10}\partial^2 T$.
The structure constants $\alpha$ and
$\beta$ appear in the three-point functions $\langle WWT \rangle$ and
$\langle WW  \Lambda^{\rm qu}\rangle$ and can be fixed either directly
from that or via the Jacobi-identities
to the values $\alpha=2$ and $\beta^{\rm qu}=32/(5c+22)$. 
In the classical $\hbar\to 0$ limit \cite{Bowcock:1991zk} the algebra becomes 
\eq{
\label{w3class}
                [L_m,L_n]&={c\over 2} {m+1\choose 3}\delta_{m+n, 0} +(m-n) L_{m+n}\,, \\[0.1cm]
               [L_m,W_n]&=(2m-n) W_{m+n}\,, \\[0.1cm]
             [W_m,W_n]&={c\over 3} {m+2\choose 5}\delta_{m+n, 0}+ {\alpha\over 60} (m-n)(2m^2+2n^2-mn-8) L_{m+n}\\
&\phantom{=}+{\beta^{\rm cl}\over 2} (m-n) \Lambda^{\rm cl}_{m+n}\, ,
}
where the commutator is meant to be a  Poisson-bracket,
i.e. $[\cdot,\cdot] := i \{\cdot,\cdot\}_{\rm PB}$.
In this limit the structure constants become $\alpha=2$,
$\beta^{\rm cl}=32/(5c)$ and the normal ordered product $\Lambda^{\rm qu}$
simplifies to  $\Lambda^{\rm cl}(z)=T(z)\cdot T(z)$,
involving just the usual product of functions.
It is this classical ${\cal W}$ algebra that will be related to an
L$_\infty$ structure in the next section.

Let us close this brief section with a couple of   remarks.
This classical ${\cal W}_3$ algebra is just the first in the series of
so-called ${\cal W}_N$ algebras which  contain $N-1$ generators of conformal
dimensions $\{2,3,\ldots,N\}$. In the context of the higher spin
AdS$_3$-CFT$_2$ duality \cite{Henneaux:2010xg,Campoleoni:2010zq,Gaberdiel:2010pz},  the  classical ${\cal W}_\infty[\mu]$
algebra played an  important role. This extremely huge non-linear algebra  can be considered as a continuous
extrapolation of the set of ${\cal W}_N$ algebra, in the sense that it
truncates as ${\cal W}_\infty[N]={\cal W}_N$. 
It is the asymptotic
symmetry algebra of Vasiliev's ${\rm hs}[\mu]$ higher spin theory.


\section{${\cal W}$ algebras and L$_{\infty}$ algebra}

In order to relate the classical ${\cal W}$ algebra to an
L$_{\infty}$ algebra, not only do we have to consider the fields
$T(z),W(z)$ but also how they transform under infinitesimal symmetry
transformations.  In this section, we first outline the general
procedure of how a classical ${\cal W}$ algebra gives rise to 
an L$_{\infty}$ structure. Second, we provide the classical ${\cal
  W}_3$ algebra 
as a concrete example.

\subsection{The general picture}
\label{sec_rela}

As mentioned   we have to define  $\delta_{\varepsilon_i}
W_j$, i.e. the infinitesimal transformation of  the fields 
$\{W_2, W_3,\ldots \}$ under the symmetry generated by
$W_i$. Here we have written $W_2=T$. 
The  L$_{\infty}$ algebra consists of two graded vector-spaces $X_0$
and $X_{-1}$.
Here, the field space $X_{-1}$ is a direct sum of the
$W_i$, thus $X_{-1} = \oplus W_i$ and the space of gauge
parameters is a direct sum of transformations with generators $W_i$,
thus $X_0 = \oplus \varepsilon_i$. 

The infinitesimal variation of the chiral fields under the symmetries
generated by $W_i$ can be determined  using 
\eq{   
\label{hannover96}
 \delta_{\varepsilon_i}  W_j(z) ={1\over 2\pi i} \oint dw
  \, \varepsilon_i(w) \,\Big[ W_i(w), W_j(z)\Big]\, ,
}
where $[\cdot,\cdot]:= i \{\cdot,\cdot\}_{\rm PB}$.
The right hand side can be generically evaluated using the form of the
OPE between quasi-primary fields \cite{Blumenhagen:1990jv} (see also
\cite{Blumenhagen:2009zz} for a more pedagogical exposition)
\eq{
           W_i(w)\, W_j(z)=\sum_k  C_{ij}{}^k\, {a_{ijk}^n\over n!} {\partial^n
             \phi_k(z)\over (w-z)^{h_i+h_j-h_k-n}}
}
with 
\eq{
        a_{ijk}^n={2h_k +n -1 \choose n}^{-1} {h_k+h_i-h_j +n -1 \choose n}\,.
}
 In a ${\cal W}$ algebra the fields $\phi_k(z)$ can 
themselves be products of the generators $W_i(z)$.
Note that the coefficients $a_{ijk}^n$ only depend on the conformal
dimensions of the fields involved and are the same in the quantum and classical
case. Only the structure constants $C_{ij}{}^k$ and the concrete form
of the product of fields receive quantum corrections. 
Thus, for the variation of the generators we obtain
\eq{    
\label{variaW}
\delta_{\varepsilon_i}  W_j(z) =
\sum_{{\footnotesize \begin{matrix} m,n\in
      \mathbb Z_0^+\\ m+n=h_{ijk}-1\end{matrix}}}
\!\!\!\! \!\!\!\! C_{ij}{}^k { a_{ijk}^n \over m!\,
    n!}  \; \partial^{m} \varepsilon_i(z)\;  \partial^n \phi_k(z)
}
with $h_{ijk}=h_i+h_j-h_k$. 

To define the higher products \eqref{higherells} of an  L$_\infty$ algebra, 
we follow \cite{Hohm:2017pnh} and  use that the closure requirement
\eq{
\label{commurel}
                      [\delta_{\varepsilon_i},\delta_{\varepsilon_j}] W_k
                      =\delta_{-\mathbf C(\varepsilon_i,\varepsilon_j, \mathbf W)} W_k
}
is equivalent to the  L$_\infty$ relations with two gauge parameters
once one identifies 
\eq{
\label{expan1}
        \delta_{\varepsilon_i} W_j =\sum_{n\ge 0}   {1\over n!}
      (-1)^{n(n-1)\over 2}\,
 \ell_{n+1}^{W_j}(\varepsilon_i, \mathbf W^n)\,
}
and
\eq{
\label{expan2}
            \mathbf C(\varepsilon_i,\varepsilon_j, \mathbf W) =\sum_l\sum_{n\ge 0} {1\over n!}
            (-1)^{{n(n-1)\over 2}}
            \;\ell^{\,\varepsilon_l}_{n+2}(\varepsilon_i,\varepsilon_j, \mathbf W^n)\,.
}
We introduced an upper index to the L$_\infty$ products 
which indicates in which $W_i$ or $\varepsilon_i$ direction the image
of the higher products is located. 
Moreover, $\mathbf C$ and $\mathbf W$ denote  elements in these vector
spaces, where  the short-hand notation $\mathbf W^n$ means that the higher product depends on $n$
generators.

The combinatorial prefactors in \eqref{expan1} and \eqref{expan2} have been determined in \cite{Hohm:2017pnh} and
guarantee that the so defined higher order products satisfy the
L$_\infty$ relations with  two gauge parameters.
Furthermore, it is shown in \cite{Hohm:2017pnh} that the L$_\infty$
relations with three gauge parameters are equivalent to a vanishing
Jacobi-identity for  three  gauge transformations
\eq{ \label{Gaugejacobiator}
\sum_{\rm cycl} \big[ \delta_{\varepsilon_i}, [  \delta_{\varepsilon_j} ,  \delta_{\varepsilon_k} ] \big] = 0 \, . 
}
Together with \eqref{commurel} this relation ensures that the gauge
transformations $\delta_{\varepsilon_i}$ form a well-defined
associative Lie algebra.

Therefore, the higher order products with one gauge parameter can be read off directly from the
expression \eqref{variaW}. Due to the appearance of products of fields
in the ${\cal W}$ algebra, in general one also gets products $\ell_n$
with $n>2$. Moreover, in \eqref{commurel} we allowed the gauge
parameter on the right hand side, $C(\varepsilon_i,\varepsilon_j,\mathbf W)$, to
depend on the fields $\mathbf W$. This structure appeared in the context of
higher spin theories in the early paper \cite{Berends:1984rq}.
In order to determine them we proceed
as follows.
Using \eqref{hannover96} and the usual Jacobi-identity for the bracket
$[\cdot,\cdot]$, one can write
\eq{
\label{werder}
      [\delta_{\varepsilon_i},\delta_{\varepsilon_j}] W_l=-{1\over 2\pi
        i}\int \sum_k C_{ij}{}^k \, {\cal P}_{ijk}(\varepsilon_i,\varepsilon_j) \, [\phi_k,W_l]
}
with universal ${\cal P}_{ijk}(\varepsilon_i,\varepsilon_j)$ that only
depend on the conformal dimensions of the fields involved
\eq{
  {\cal
    P}_{ijk}(\varepsilon_i,\varepsilon_j)=\sum_{{\footnotesize \begin{matrix} r,s\in
      \mathbb Z_0^+\\ r+s=h_{ijk}-1\end{matrix}}}
  \!\!\!\! \!\!\!\! \kappa_{ijk}^{rs}\; \partial^r\varepsilon_i \;\partial^s \varepsilon_j
}
with 
\eq{
                    \kappa_{ijk}^{rs}={(-1)^r (2h_k-1)!\over r! s!
  (h_i+h_j+h_k-2)!} \prod_{t=0}^{s-1}  (2h_i -2-r-t) \prod_{u=0}^{r-1}  (2h_j -2-s-u) \,.
}
Using the Leibniz-rule for $[\phi_k,W_l]$ in \eqref{werder}, one can
read off   
\eq{
\label{higherc1}
       \mathbf  C(\varepsilon_i,\varepsilon_j,\mathbf W)= \sum_l \sum_k C_{ij}{}^k \,  {\cal
           P}_{ijk}(\varepsilon_i,\varepsilon_j) \,\partial_l \phi_k  \in X_0\,.
}
Using \eqref{expan2} and expanding the right-hand side, one can read
off the higher order L$_\infty$ products with two gauge parameters.
If we have fixed the structure constants of the initial ${\cal W}$-algebra
such that the Jacobi-identity for the Poisson bracket is satisfied
then
by construction it is guaranteed that the L$_\infty$ relations are
satisfied. However, as we will see, one can 
initially  leave some of the
structure constants open and  fix them by the 
L$_\infty$ relations. In the CFT context this  is called the
bootstrap approach.


\subsection{The classical ${\cal W}_3$ algebra}

Since the general formalism developed in the previous section might
appear quite abstract,  let us now exemplify all this explicitly for the classical  ${\cal W}_3$ algebra.


\subsubsection{L$_\infty$ products with one gauge parameter}
Using the algebra \eqref{w3class} and the general relations \eqref{variaW}, \eqref{expan1},
one can derive the infinitesimal variations and read off the higher
products with one gauge parameter
\begin{align} \label{var1}
 \delta_{\varepsilon} T&=\underbrace{{c\over 12} \, \partial^3 \varepsilon}_{\ell^{T}_1(\varepsilon)}
        + \underbrace{(2\, \partial \varepsilon\, T +  \varepsilon \, \partial T)}_{\ell^{T}_2(\varepsilon,T)} \,,\nonumber \\[0.2cm]
       \delta_{\varepsilon} W&=
        \underbrace{ (3\, \partial \varepsilon\, W +  \varepsilon \, \partial W)}_{\ell^{W}_2(\varepsilon,W)} 
        \intertext{and} \nonumber 
  \delta_{\eta} W&=\underbrace{{c\over 360} \, \partial^5
          \eta}_{\ell^{W}_1 (\eta)}
        + \underbrace{\alpha\Big({1\over 6} \, \partial^3 \eta\, T +
          {1\over 4} \, \partial^2
          \eta\, \partial T+ {3\over 20}\,  \partial
          \eta\, \partial^2 T+{1\over 30}\, 
          \eta\, \partial^3 T\Big)}_{\ell^{W}_2(\eta,T)}
        \\[0.1cm]
        &\phantom{=}+\underbrace{\beta \Big(\partial\eta\, (TT) +{1\over 2}
        \eta\, \partial (TT)\Big)}_{-{1\over 2}\ell^{W}_3(\eta,T,T)}\,,\\[0.3cm]
       \delta_{\eta} T&=
        \underbrace{ (3\,\partial \eta\, W +  2\, \eta \, \partial W)}_{\ell_2^{T}(\eta,W)}\, ,  \nonumber       
\label{var2}
\end{align}
where here and in the following we use the
shorter notation where an $\varepsilon$ denotes an $\varepsilon_T$ and an $\eta$ always denotes an $\varepsilon_W$.
In the following  we will first  leave the two structure constants  $\alpha=C_{WW}^T$ and
$\beta=C_{WW}^{\Lambda}$ undetermined. We will see that the L$_\infty$
relations indeed fix them to their expected values $\alpha=2$ and 
$\beta=32/(5c)$. It is amusing that the central terms in the ${\cal
  W}$-algebra are related to the maps $\ell_1:X_0\to X_{-1}$.


\subsubsection{L$_\infty$ products with two gauge parameters}

By taking \eqref{higherc1} as an ansatz  and comparing it with \eqref{expan2}, we can read off the L$_\infty$ products with two gauge parameters. We find
\begin{align}
\mathbf C(\varepsilon_1,\varepsilon_2,\mathbf W)&=\varepsilon_1 \partial\varepsilon_2-\partial \varepsilon_1\, \varepsilon_2 =: \ell_2^{\,\varepsilon}(\varepsilon_1,\varepsilon_2)\, ,\\[0.2cm]
\mathbf C(\, \varepsilon \, , \, \eta \, , \mathbf W)&=\varepsilon \, \partial\eta-2\, \partial\varepsilon\, \eta
 =: \ell^{\,\eta}_2(\varepsilon,\eta)
\intertext{and a non-trivial higher order correction in}
\label{WWfield}
    \mathbf  C(\eta_1,\eta_2, \mathbf  W) & =\ell^{\,\varepsilon}_2(\eta_1,\eta_2)+\ell^{\,\varepsilon}_3(\eta_1,\eta_2,T)
\end{align}
with 
\begin{align}
\label{sonne}
    \ell^{\,\varepsilon}_2(\eta_1,\eta_2) &=\alpha \left(  {1\over
        30} \eta_1\,\partial^3 \eta_2  -{1\over
        30} \partial^3\eta_1\,\eta_2
+{1\over    20} \partial^2\eta_1\,\partial\eta_2-{1\over
        20} \partial\eta_1\,\partial^2\eta_2\right)
\end{align}
and 
\begin{align}
\label{mond}
\ell^{\,\varepsilon}_3(\eta_1,\eta_2,T)&=\beta\, T\left( \eta_1 \partial\eta_2-\partial \eta_1\, \eta_2\right)\,.
\end{align}
Please note the highly non-trivial form of \eqref{sonne} and the $T$
dependence in \eqref{mond}.


\subsubsection{L$_\infty$ relations with two gauge parameters and closure}
The following section is dedicated to the equivalence of the closure condition
\eq{
\label{commurel2}
                      [\delta_{\varepsilon_i},\delta_{\varepsilon_j}] W_k
                      =\delta_{-\mathbf C(\varepsilon_i,\varepsilon_j, \mathbf W)} W_k\,  
}
and the L$_\infty$ relations \eqref{ininftyrel} and
\eqref{ininftyrel2}. 
For every combination $(ij,k)$ we will explicitly state the
corresponding L$_\infty$ relations. 
\begin{itemize}
\item \textbf{(TT,T):} The closure condition \eqref{commurel2} is equivalent to the L$_\infty$ relations
\begin{eqnarray}
             \ell^{T}_1 \big
(\ell^{\,\varepsilon}_2(\varepsilon_1,\varepsilon_2)\big
)   & =& \ell^{T}_2\big(\ell^{T}_1(\varepsilon_1),
                 \varepsilon_2\big) + \ell^{T}_2\big(\varepsilon_1,\ell^{T}_1(\varepsilon_2)\big) \,,\\[0.2cm]
                    0 &=&\ell^{T}_2\big(\ell^{\,\varepsilon}_2(\varepsilon_1,\varepsilon_2),T\big)+\ell^{T}_2\big(\ell^{T}_2(\varepsilon_2,T),
                 \varepsilon_1\big) + \ell^{T}_2\big(\ell^{T}_2(T,\varepsilon_1),\varepsilon_2\big) \,.\nonumber
\end{eqnarray}

\item \textbf{(TT,W):} Due to  $\ell_1^{W}(\varepsilon)=0$ the closure condition corresponds to the non-trivial relation 
\eq{
    0=\ell^{W}_2\big(\ell^{\,\varepsilon}_2(\varepsilon_1,\varepsilon_2),W\big)+\ell^{W}_2\big(\ell^{W}_2(\varepsilon_2,W),
                 \varepsilon_1\big) + \ell^{W}_2\big(\ell^{W}_2(W,\varepsilon_1),\varepsilon_2\big) \, .
}

\item \textbf{(TW,T):}
One gets the single non-trivial relation
\eq{
                 0=  \ell^{T}_2\big(\ell^{\,\eta}_2(\varepsilon,\eta),W\big)+
 \ell^{T}_2\big(\ell^{T}_2(\eta ,W),\varepsilon\big) +
\ell^{T}_2\big(\ell^{W}_2(W,\varepsilon),\eta\big) \,.
}

\item {\bf (TW,W):} In this case the closure condition also involves the products $(TT)$ and therefore for the first time implies higher order
L$_\infty$ relations. One finds
\eq{
\label{longrelats}
             \ell^{W}_1\big(\ell^{\,\eta}_2(\varepsilon,\eta)\big)     & =  \ell^{W}_2\big(\ell^{T}_1(\varepsilon),
                 \eta\big) +
                 \ell^{W}_2\big(\varepsilon, \ell^{W}_1(\eta)\big) \,
}
which holds only for $\alpha = 2$, and 
\eq{             
                   0 & =\phantom{+}\ell^{W}_2\big(\ell^{\,\eta}_2(\varepsilon,\eta),T\big)+\ell^{W}_2\big(\ell^{W}_2(\eta,T),
                 \varepsilon\big) +           \ell^{W}_2\big(\ell^{T}_2(T,\varepsilon),\eta\big)
                 \\[0.1cm]     &
                 \phantom{=} \; +\ell^{W}_3\big(\ell^{T}_1(\varepsilon),\eta,T\big) \, 
}
that vanishes only for $16\alpha = 5c \beta$.  Finally, one also has the
non-trivial ${\cal J}_4$ relation
\eq{
\label{sterne}
        0 & =\ell^{W}_3\big( \ell^{\,\eta}_2(\varepsilon,\eta),T,T\big)-2 \, \ell^{W}_3\big(\ell^{T}_2(\varepsilon,T),\eta,T\big)
-\ell^{W}_2\big(\varepsilon,\ell^{W}_3(\eta,T,T) \big) \,  
} \noindent
that is zero for any choice of $\alpha$ and $\beta$. 
Together these are precisely the values familiar from the classical ${\cal
  W}_3$ algebra. Note that here we have only presented  those terms from the general L$_\infty$
relation that are non-vanishing. The combinatorial factor of two in
\eqref{sterne} is due to the exchange of the two $T$-fields.

\item {\bf (WW,T):} Here another new aspect appears, namely that the closure condition requires a higher order correction to $\mathbf C(\eta_1,\eta_2,\mathbf  W)$ \eqref{WWfield}. The non-trivial parts in the  L$_\infty$ relations now read
\eq{
\label{longrelatsb}
            \ell^{T}_1\big(\ell^{\,\varepsilon}_2(\eta_1,\eta_2)\big)    & =\ell^{T}_2\big(\ell^{W}_1(\eta_1),
                 \eta_2\big) +\ell^{T}_2\big(\eta_1,\ell^{W}_1(\eta_2) \big)\,,
                  \\[0.2cm]
              0 &= \ell^{T}_1\big(\ell^{\,\varepsilon}_3(\eta_1,\eta_2,T)\big)+    \ell^{T}_2\big(\ell^{\,\varepsilon}_2(\eta_1,\eta_2),T\big)
 \\ &\phantom{aaaaaa}+\ell^{T}_2\big(\ell^{W}_2(\eta_2,T),\eta_1\big) 
      + \ell^{T}_2\big(\ell^{W}_2(T,\eta_1),\eta_2\big)  \,,\\[0.2cm]
      0 & =      -2
      \ell^{T}_2\big(\ell^{\,\varepsilon}_3(\eta_1,\eta_2,T),T\big) \\
	& \phantom{aaaaaa}      
      +\ell^{T}_2\big(\eta_2,\ell^{W}_3(\eta_1,T,T)
     \big ) -
     \ell^{T}_2\big(\eta_1,\ell^{W}_3(\eta_2,T,T) \big) \, . 
}
Again, the ${\cal J}_2$ relation is satisfied for $\alpha=2$ and the ${\cal J}_3$ relation requires $16\alpha = 5c \beta$ to hold.

\item 
{\bf  (WW,W):} For this final closure condition, we find the
equivalence to the L$_\infty$ relations
\begin{eqnarray}
\label{longrelatsc}
          && 0=\ell^{W}_2\big(\ell^{\,\varepsilon}_2(\eta_1,\eta_2),W\big)
 +\ell^{W}_2\big(\ell^{T}_2(\eta_2,W),\eta_1\big) 
      + \ell^{W}_2\big(\ell^{T}_2(W,\eta_1),\eta_2\big)\,,\\[0.2cm]
     &&    0=     \ell^{W}_2\big(\ell^{\,\varepsilon}_3(\eta_1,\eta_2,T),W\big)+\ell^{W}_3\big(\ell^{T}_2(\eta_1,W),\eta_2,T\big)
              +\ell^{W}_3\big(\eta_1,\ell^{T}_2(\eta_2,W),T\big)  \nonumber
\end{eqnarray}
that are both independent of $\alpha$ and $\beta$.

\end{itemize}

\vspace{0.2cm}
\noindent
One can also show that the above relations cover  all non-trivial
L$_\infty$ relations with two gauge parameters, so that
we have  explicitly checked their equivalence to  the closure conditions.


\subsubsection{L$_\infty$ relations with three gauge parameters and
  the Jacobi-identity}

It only remains to check the L$_\infty$ with three gauge
parameters. Recall that they are supposed to be  equivalent to the
Jacobi-identity \eqref{Gaugejacobiator} of three
gauge transformations, i.e.
\eq{ 
\label{Gaugejacobiator2}
\sum_{\rm cycl} \big[ \delta_{\varepsilon_i}, [  \delta_{\varepsilon_j} ,  \delta_{\varepsilon_k} ] \big] = 0 \, . 
}
Again we  list the L$_\infty$ relations that ensure this relation for a given combination of $(ijk)$.
\begin{itemize}
\item {\bf (TTT):}
\eq{
0=\ell^{\,\varepsilon}_2\big(\ell^{\,\varepsilon}_2(\varepsilon_1,\varepsilon_2),\varepsilon_3\big)+\ell^{\,\varepsilon}_2\big(\ell^{\,\varepsilon}_2(\varepsilon_3,\varepsilon_1),\varepsilon_2\big)+\ell^{\,\varepsilon}_2\big(\ell^{\,\varepsilon}_2(\varepsilon_2,\varepsilon_3),\varepsilon_1\big) \nonumber
}

\item {\bf (TTW):}
\eq{
0=\ell^{\,\eta}_2\big(\ell_2^{\,\varepsilon}(\varepsilon_1,\varepsilon_2),\eta\big)+\ell^{\,\eta}_2\big(\ell_2^{\,\eta}(\eta,\varepsilon_1),\varepsilon_2\big)+\ell^{\,\eta}_2\big(\ell_2^{\,\eta}(\varepsilon_2,\eta),\varepsilon_1\big) \nonumber
}

\item {\bf (WWT):}
\begin{align*}
&0=\ell^{\,\varepsilon}_2\big(\ell_2^{\,\varepsilon}(\eta_1,\eta_2),\varepsilon\big)+\ell^{\,\varepsilon}_2\big(\ell_2^{\,\eta}(\varepsilon,\eta_1),\eta_2\big)+\ell^{\,\varepsilon}_2\big(\ell_2^{\,\eta}(\eta_2,\varepsilon),\eta_1\big)\\[0.1cm]
	&\phantom{0=}+\ell^{\,\varepsilon}_3\big(\eta_1,\eta_2,\ell^{T}_1(\varepsilon)\big)\,,
\\[0.3cm]
& 0=-\ell^{\,\varepsilon}_2\big(\ell^{\,\varepsilon}_3(\eta_1,\eta_2,T),\varepsilon\big)+\ell^{\,\varepsilon}_3\big(\ell^{\,\eta}_2(\eta_1,\varepsilon),\eta_2,T\big)\,,\\[0.1cm]
& \phantom{0=} -\ell^{\,\varepsilon}_3\big(\ell^{\,\eta}_2(\eta_2,\varepsilon),\eta_1,T\big)+\ell^{\,\varepsilon}_3\big(\ell^{T}_2(T,\varepsilon),\eta_1,\eta_2\big) \,.
\end{align*}
The first ${\cal J}_3$-type relation requires $16 \alpha = 5c \beta$
to hold  and shows that the two-product $\ell_2$ violates the Jacobi-identity.

\item {\bf (WWW):}
\begin{align*}
0&=\ell^{\,\eta}_2\big(\ell_2^{\,\varepsilon}(\eta_1,\eta_2),\eta_3\big)+\ell^{\,\eta}_2\big(\ell_2^{\,\varepsilon}(\eta_3,\eta_1),\eta_2\big)+\ell^{\,\eta}_2\big(\ell_2^{\,\varepsilon}(\eta_2,\eta_3),\eta_1\big) \,,  \\[0.2cm]
0&=\ell^{\,\eta}_2\big(\ell_3^{\,\varepsilon}(\eta_1,\eta_2,T),\eta_3\big)+\ell^{\,\eta}_2\big(\ell_3^{\,\varepsilon}(\eta_3,\eta_1,T),\eta_2\big)+\ell^{\,\eta}_2\big(\ell_3^{\,\varepsilon}(\eta_2,\eta_3,T),\eta_1\big) \, , \\[0.2cm] 
0&=\ell^{\,\varepsilon}_3\big(\ell_2^{T}(\eta_1,W),\eta_2,\eta_3\big)+\ell^{\,\varepsilon}_3\big(\ell_2^{T}(\eta_2,W),\eta_3,\eta_1\big)+\ell^{\,\varepsilon}_3\big(\ell_2^{T}(\eta_3,W),\eta_1,\eta_2\big)\,.
\end{align*}
\end{itemize}
Again the listed relations cover all non-trivial L$_\infty$ relations
with three gauge parameters. In particular, all relations of type
${\cal J}_5=0$ are trivially satisfied for the ${\cal W}_3$ algebra.


\subsubsection{Summary}

Looking back one realizes  that the constraint $\alpha = 2$ arose
solely from the relations ${\cal J}_2 = 0$, while $16 \alpha = 5 c
\beta$ was needed for ${\cal J}_3 =0$ to hold. The other non-trivial
L$_\infty$ relations of type  ${\cal J}_4 = 0$ were satisfied
regardless of the  particular value $\alpha$ and $\beta$.

Therefore for the known classical values $\alpha = 2$ and $\beta = 32/(5c)$ all higher order relations of the L$_\infty$ algebra are
fulfilled and we have shown that the classical ${\cal W}_3$ algebra
induces  a highly non-trivial  L$_\infty$ algebra.
Since the concrete form of the ${\cal W}_3$ algebra was known before,
it looks like that we have just checked what we argued must be true
on general grounds in section \ref{sec_rela}. However, from the
L$_\infty$ point of view, the general conformal field theory structure
provides a concrete recipe to read off higher order products that satisfy 
all L$_\infty$ relations.

Moreover, as we actually did, one  can  turn the logic around and determine
the unknown structure constants and relations among them 
 from the L$_\infty$ relations.
In other words, with general input from CFT, one can also use the L$_\infty$ structure to {\it bootstrap}
the structure constants in the extended classical conformal symmetry algebras.


\section{Conclusions}

The purpose of this letter was to point out  that classical ${\cal
  W}$ algebras do naturally define an  L$_\infty$ algebra whose
objects are concentrated in $X_0$ and $X_{-1}$. We provided a
general prescription of how to read off the higher order L$_\infty$  products from 
the conformal field theory structure. Moreover,  we argued that  the 
Poisson algebra relations for the ${\cal
  W}$ algebra are equivalent to the non-trivial
higher order   L$_\infty$ relation for two and three gauge parameters. 
As an application, we explicitly worked out the example of a ${\cal
  W}_3$ algebra.
In that case, the highest
appearing products are of order three and as  a consequence only
a finite number of higher order relations are non-trivial.
Turning the logic around,  the structure constants of the  ${\cal
  W}_3$ could also be  bootstrapped  by the L$_\infty$ relations.

 It should be clear how this generalizes to ${\cal W}$ algebras with
 more generators.  For instance, the  ${\cal W}_N$ algebra contains
generators of conformal dimension $\{2,3,\ldots,N\}$. The highest
order product of fields is $(T^{N-1})$ and appears  in
$\delta_{\varepsilon_N} W_N$. Therefore, the expectation is that this
also defines an L$_\infty$ algebra with
the highest higher order product being $\ell^{W_N}_{N}(\varepsilon_N, T^{N-1})$.
Of course, in this case  higher order relations up to ${\cal J}_{2N-1}$
  can become non-trivial. 
Therefore, the classical ${\cal W}_N$ algebras provide  an infinite
set of highly non-trivial  L$_\infty$ algebras.
One could also consider classical super ${\cal W}$ algebras, which
will
correspond to super L$_\infty$ algebras.

Coming back to  the AdS$_3$-CFT$_2$
duality, one might suspect that the classical ${\cal W}_\infty[\mu]$ algebra is related to an  L$_\infty$ algebra with 
an infinite number of higher products satisfying  an infinite number
of higher order relations. As mentioned, for positive integer $\mu$ one has the truncation
${\cal W}_\infty [N]={\cal W}_N$ which will also hold
  for the L$_\infty$ algebra.

So far we were discussing just the classical ${\cal W}$ symmetry and
one could wonder whether also the quantum case admits a description
in terms of an L$_\infty$ algebra or a quantum version of it. This
questions seems to be fairly non-trivial, as the procedure laid out in
section \ref{sec_rela} does not go through straightforwardly. In the
quantum case all products of fields  need to be normal ordered and it
is not clear what a field dependent gauge parameter should mean.
We leave this  interesting question for future research. 

Finally, one might ask whether one can generalize this  L$_\infty$
structure beyond
\begin{itemize}
\item{the two vector spaces $X_0$ and $X_{-1}$\,,}
\item{the chiral sector of a 2D CFT\,,}
\item{two-dimensions\,.}
\end{itemize}


\noindent
\emph{Acknowledgments:} We are very grateful to Andreas Deser  for
sharing part of his knowledge on  L$_\infty$ algebras with us. We also
thank Jim Stasheff for useful comments about the first version of this article.


\clearpage
\bibliography{references}  
\bibliographystyle{utphys}


\end{document}